\documentclass[manuscript]{aastex}

\usepackage{aas_macros,url,natbib,amssymb,amsmath,CJK}

\shorttitle{C/Siding Spring: Breathtaker or Nightmare?}
\shortauthors{Ye \& Hui}

\begin{document}
\begin{CJK*}{UTF8}{gbsn}

\title{An Early Look of Comet C/2013 A1 (Siding Spring): Breathtaker or Nightmare?}

\author{Quan-Zhi Ye (叶泉志)}
\affil{Department of Physics and Astronomy, The University of Western Ontario, London, Ontario, N6A 3K7, Canada}
\email{qye22@uwo.ca}

\and

\author{Man-To Hui (许文韬)}
\affil{Nhut Thung Pau Observatory, Guangzhou, China}

\begin{abstract}
The dynamically new comet, C/2013 A1 (Siding Spring), is to make a close approach to Mars on 2014 October 19 at 18:30~UT at a distance of $40\pm1$ Martian radius. Such extremely rare event offers a precious opportunity for the spacecrafts on Mars to closely study a dynamically new comet itself as well as the planet-comet interaction. Meanwhile, the high speed meteoroids released from C/Siding Spring also pose a threat to physically damage the spacecrafts. Here we present our observations and modeling results of C/Siding Spring to characterize the comet and assess the risk posed to the spacecrafts on Mars. We find that the optical tail of C/Siding Spring is dominated by larger particles at the time of the observation. Synchrone simulation suggests that the comet was already active in late 2012 when it was more than 7~AU from the Sun. By parameterizing the dust activity with a semi-analytic model, we find that the ejection speed of C/Siding Spring is comparable to comets such as the target of the Rosetta 
mission, 67P/Churyumov-Gerasimenko. Under nominal situation, the simulated dust cone will miss the planet by about 20 Martian radius. At the extreme ends of uncertainties, the simulated dust cone will engulf Mars, but the meteoric influx at Mars is still comparable to the nominal sporadic influx, seemly indicating that intense and enduring meteoroid bombardment due to C/Siding Spring is unlikely. Further simulation also suggests that gravitational disruption of the dust tail may be significant enough to be observable at Earth.
\end{abstract}


\section{Introduction}

Near-Earth Objects (NEOs) play an important role in shaping the geological histories of terrestrial planets. Recent studies have shown NEO impacts are common in inner solar system \citep[c.f.][]{str05}. For other terrestrial planets, it has been suggested that the impact flux is comparable to the near-Earth environment \citep{feu11}. Although over 99\% of the impactors are asteroids \citep{yeo13}, comets are generally of special interests, as they carry significant amount of volatile and organic material, which is life-essential. On Earth, kilometer-sized cometary impacts occur every $\sim10^{8}$~yr \citep{sto03}.

On the other hand, close comet-planet approach is also significant in terms of the accretion of water and organic materials on the planet: comets eject a large amount of material into the vicinity of their nuclei, and they may still influence the planet without a direct impact. Although approaches are more common than impacts, it is still too rare for us to observe and study a real case: the closest cometary approach to the Earth since the establishment of modern science was D/1770 L1 (Lexell), which missed the Earth by $\sim 356$ Earth radius. From the impact rates, we estimate that close approach within 25 Earth radius with kilometer-sized comets occurs once every $\sim10^5$~yr. This is equivalent to the frequency of cometary approach within 50 Martian radius to Mars assuming that the cometary impact flux (like the total impact flux) is comparable between Earth and Mars.

\begin{figure*}
\includegraphics[width=\textwidth]{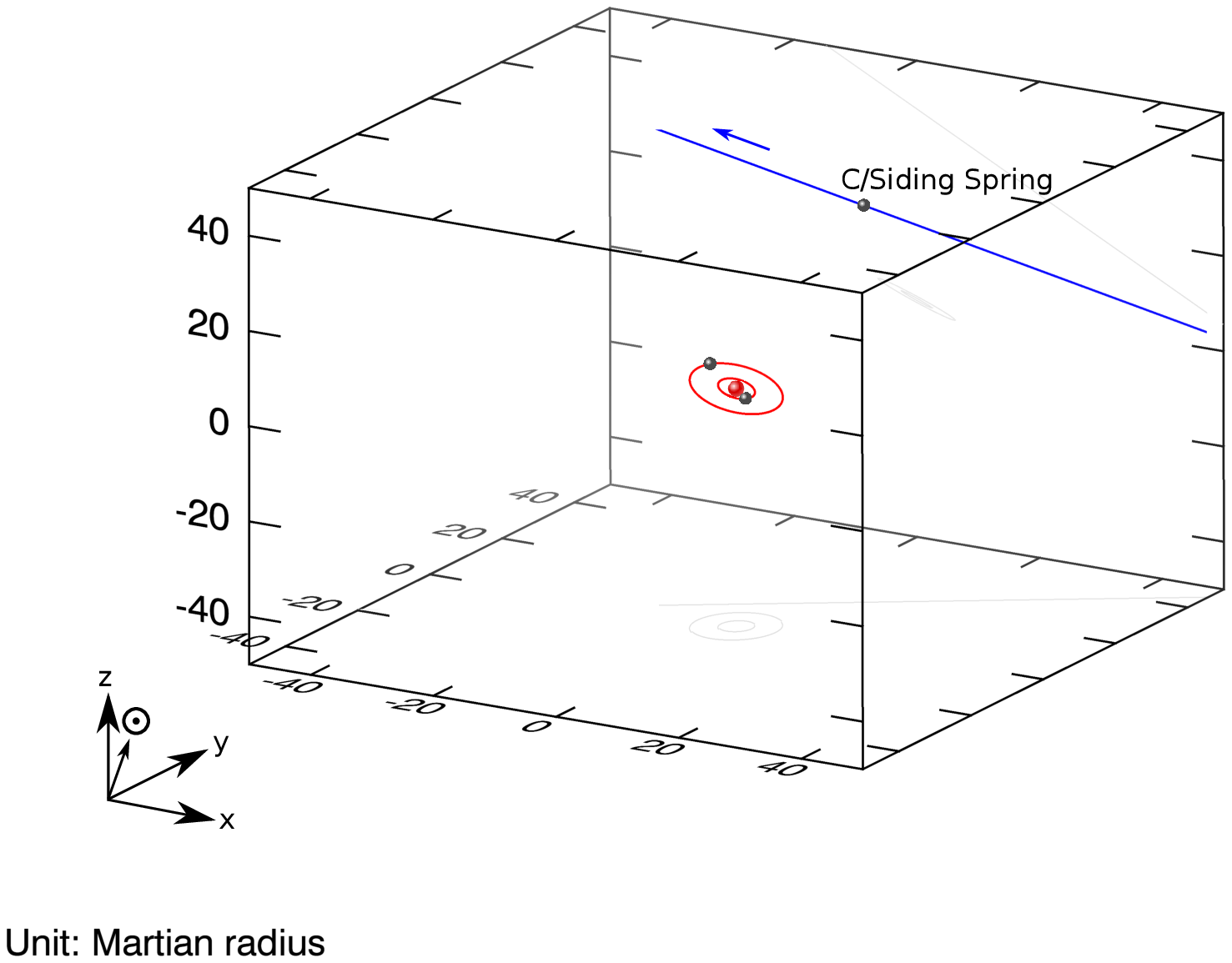}
\caption{The close approach of C/Siding Spring to Mars. The plot is constructed in J2000 ecliptic frame with 1 unit equals to 1 Martian radius. The two objects orbiting Mars (center object in red) are the two Martian moons, Phobos (inner) and Deimos (outer). The sizes of the objects are not to scale.}
\label{fig-a1orb}
\end{figure*}

\begin{figure*}
\includegraphics[width=\textwidth]{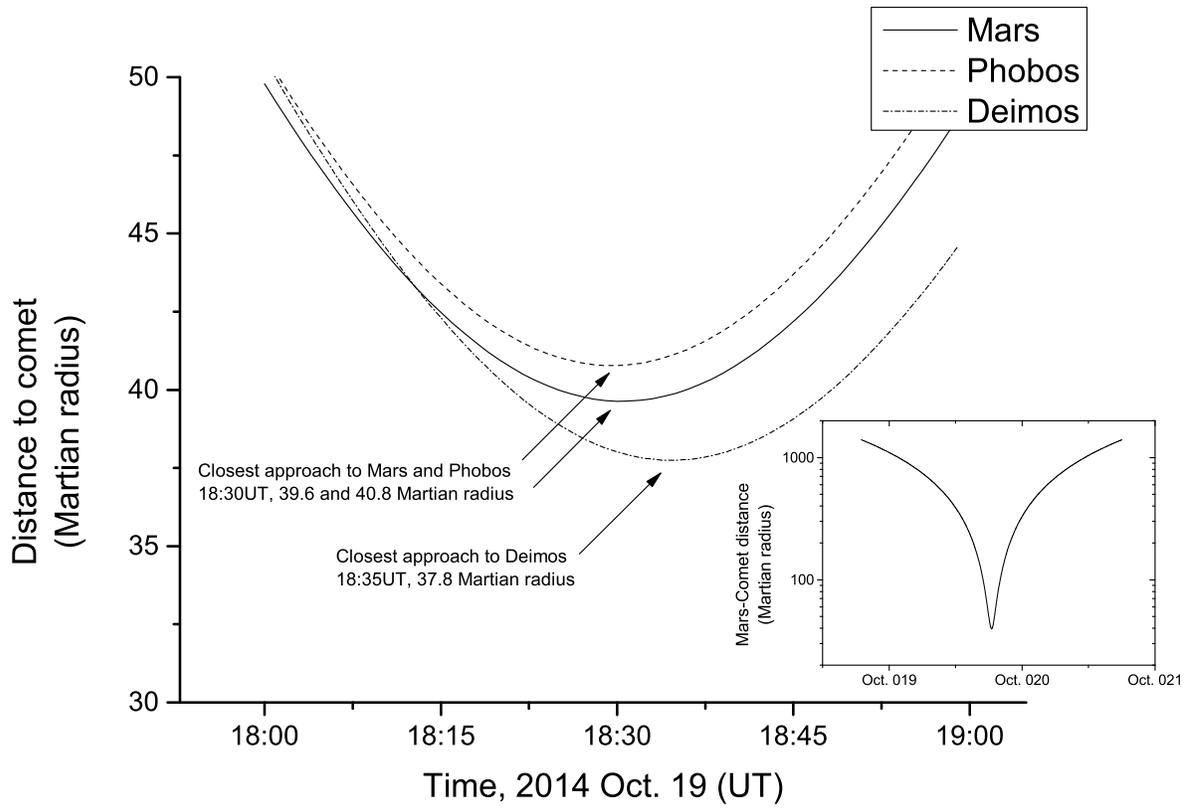}
\caption{The distance between C/Siding Spring and Mars, Phobos and Deimos within 30~min. of the closest approach (small graph is the Mars-comet distance within 1~day of the closest approach). The miss distances have an uncertainty of $\sim1$ Martian radius.}
\label{fig-dist}
\end{figure*}

Yet this is what would happen later this year: a dynamically new comet, C/2013 A1 (Siding Spring), is to miss Mars by $\sim 40$ Martian radius at 2014 Oct. 19.8 (UT) (Figure~\ref{fig-a1orb} and \ref{fig-dist}). C/Siding Spring was discovered on 2013 Jan. 3 at a heliocentric distance of 7.2~AU; subsequent follow-up observations revealed a 10'' coma which indicated distinct cometary activity at such a large heliocentric distance \citep{mcn13}. As of 2014 Feb. 1, the comet is determined to be in a hyperbolic orbit, with $e=1.0006$; the current estimated miss distance between C/Siding Spring and Mars is about $40\pm1$ Martian radii or $135600\pm6000$~km\footnote{Update numbers can be found at \url{http://ssd.jpl.nasa.gov/sbdb.cgi?sstr=2013 A1}.}.

Dynamically new comets are constrained on loosely bounded or unbounded orbits, and are thought to originate from the outer region of solar system, namely the Oort cloud. Due to the fact that they have had nil access to the inner solar system, they preserve valuable and unique information about the pre-solar nebula. However, comparing to the periodical comets, which usually return to the inner solar system on a frequent and predictable basis, the dynamically new comets are difficult to investigate due to their small number and limited opportunity to study individual objects (generally only once). The most productive method to study comets -- in-situ exploration -- is currently very difficult to be used on dynamically new comets, due to very short lead-time available for preparing and operating such missions.

As such, the close approach offers an unprecedented and extremely rare opportunity to directly study how material may be transferred from comets to terrestrial planets as well as the dynamically new comet itself. Currently there are three operational orbiters (Mars Reconnaissance Orbiter, Mars Odyssey and Mars Express) and two operational rovers (Opportunity and Curiosity) on Mars; in addition, two orbiters (Mars Atmosphere and Volatile Evolution or MAVEN, and Mars Orbiter Mission or MOM) will arrive $\sim 1$ month before C/Siding Spring's closest approach. The fleet will have front seats for this event; however, the small miss distance of the encounter also means that they may pass inside the dust coma/tail of C/Siding Spring. While the Martian atmosphere will shield incoming dust particles (meteoroids) for the two rovers, the five orbiters will be at risk of bombardment of dust particles originated from the comet. Cometary dusts pose a significant threat of causing physical damage to the 
spacecrafts \citep[c.f.][]{ahe08}. Additionally, meteoroids originated from C/Siding Spring have higher kinetic energy than nominal sporadic (background) meteoroids, as the relative speed between the comet and Mars is twice as high as the latter. Early studies of C/Siding Spring before its close visit to Mars will be essential in the sense of monitoring the evolution of the comet and helping assess the risk posed to the spacecrafts.

Here we present our observations and modeling effort of C/Siding Spring in the hope to characterize the physical properties of the comet. We will first discuss our observations and their significance for constraining the particle size distribution (PSD) of the comet, then we will present the semi-analytic model that will be used to match the observation and parameterize the cometary dust activity. Eventually, we will use the best-matched parameters to investigate the fluency of cometary dust particles experienced by Mars (and anything in the proximity) during the close encounter.

\section{Observations and Initial Interpretation}

\subsection{Planning and Conducting the Observations}

After being released by the parent body, different sizes of dust particles follow different Keplerian trajectories as they feel different radiation pressure and gravity from the Sun. The ratio of later two quantities is defined as $\beta$, of which $\beta \propto (\rho r)^{-1}$ where $\rho$ is particle density and $r$ is size. If we ignore the initial velocity of the particles with respect to the nucleus and compute a large number of particles with different $\beta$ released at different times, we can produce the so-called syndyne-synchrone diagram as defined by \citet{fin68}.

\begin{figure}
\includegraphics[width=0.5\textwidth]{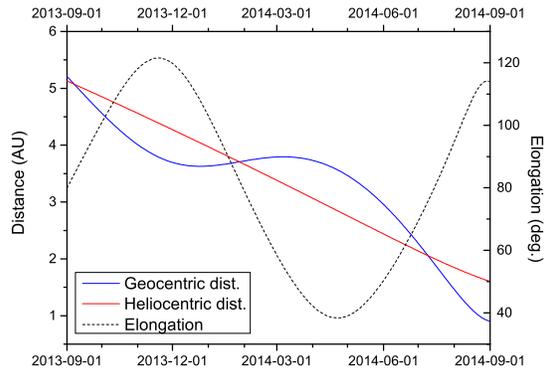}
\caption{Observation circumstances of C/Siding Spring from September 2013 to August 2014.}
\label{fig-obs-geom}
\end{figure}

\begin{figure*}
\includegraphics[width=\textwidth]{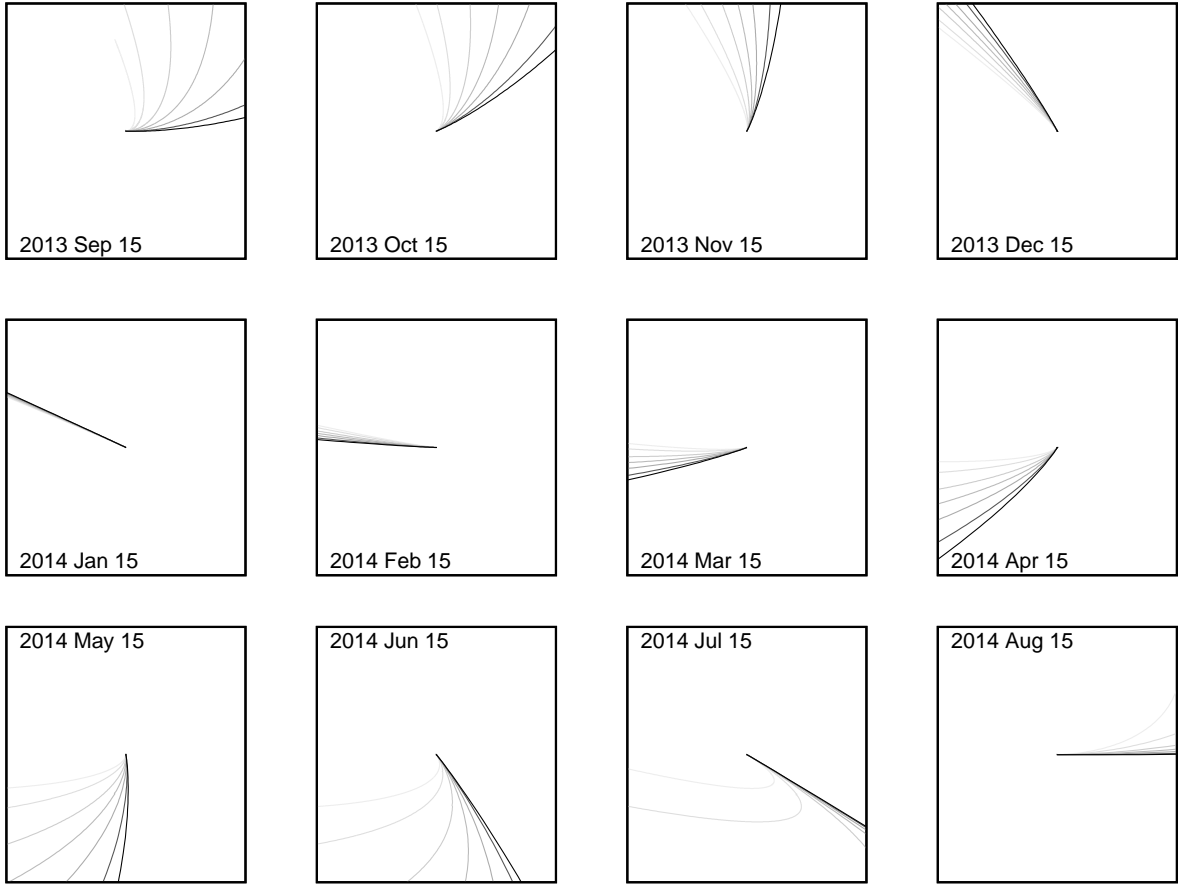}
\caption{Syndyne simulations of C/Siding Spring from September 2013 to August 2014. $\beta$ ranges from 0.01 (light grey) to 1 (black). Each diagram is $10' \times 10'$ in size, north is up and east is left.}
\label{fig-synpanel}
\end{figure*}

Occasionally, the Earth-comet geometry is favorable to so that different syndyne curves (i.e. equal-$\beta$ curves) are well separated from each other as seen by the observer, which allows us to qualitatively constrain the PSD of the comet. The occurrence of such geometry depends on the orientation of the orbital plane of the comet. We find out that for C/Siding Spring, such geometry would occur in September to November 2013, while the solar elongation of the comet is adequate for optical observations (Figure~\ref{fig-obs-geom} and \ref{fig-synpanel}). The next ``slots'' will occur in June 2014 and September 2014, but they are either suffered from small solar elongation or being too close to the encounter event.

\begin{figure*}
\includegraphics[width=\textwidth]{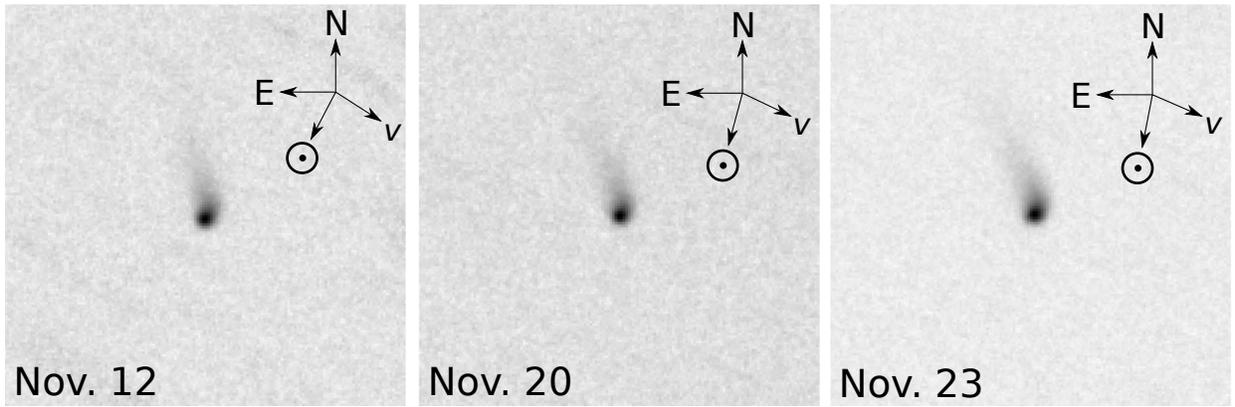}
\caption{Master frames of the observations on 2013 Nov. 12, Nov. 20, and Nov. 23. Each image is 6'$\times$4'.}
\label{fig-img}
\end{figure*}

We conduct broad-band observations with a 0.18-m f/7 refraction telescope and 4k$\times$3k CCD camera (pixel size 1.5'') at Jade Scope Observatory near Siding Spring, Australia ($149^{\circ}12'$~E, $31^{\circ}17'$~S), on November 12, 20 and 23, 2013 (details summarized in Table~\ref{tbl-obs}). Observations are unfiltered and the CCD sensor is most sensitive at $\sim 500$~nm. The raw frames are then subtracted by dark and bias fields and divided by flat fields. After initial reduction, the frames are registered with the NOMAD catalog \citep{zac04} so that they can be combined and stacked following the motion of C/Siding Spring. Eventually, we end up with three ``master'' images from each night(Figure~\ref{fig-img}).

\subsection{Is C/Siding Spring Rich in Big Particles?}

\begin{figure*}
\includegraphics[width=0.5\textwidth]{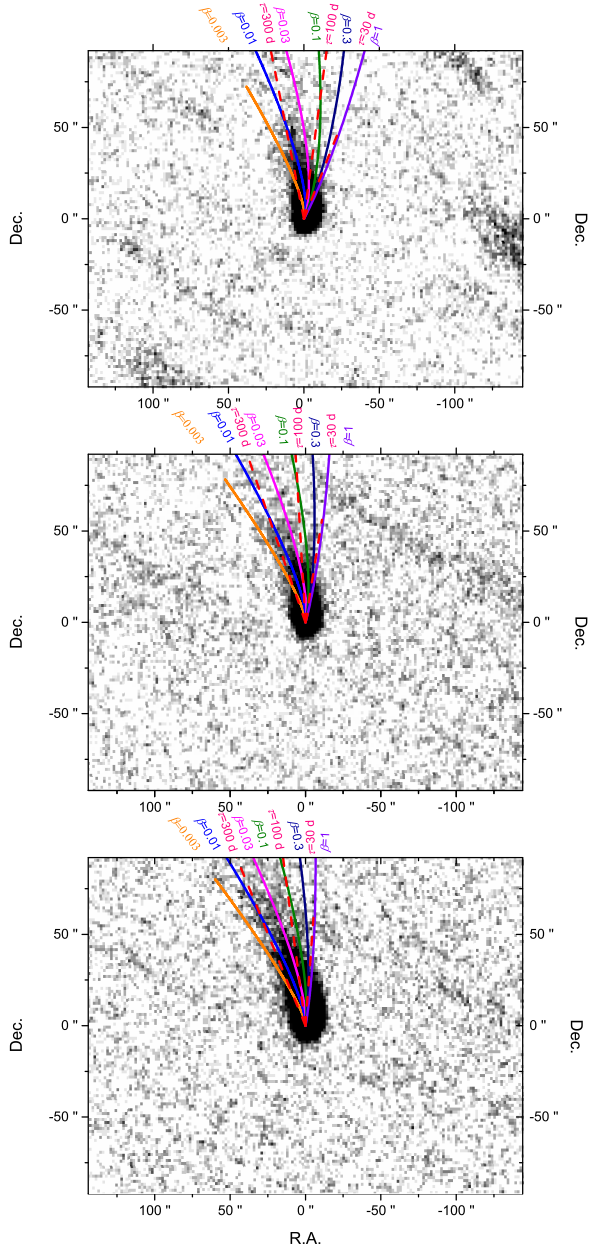}
\caption{Syndyne-synchrone computation for the master frames. North is up and east is left.}
\label{fig-syn}
\end{figure*}

We compute the syndyne-synchrone curves for the three master frames (Figure~\ref{fig-syn}) and immediately notice that C/Siding Spring's tail is skewed to smaller $\beta$ values (i.e. larger particles), dominantly at the order of 0.01. If we use $\beta=5.74\times10^{-4}/(\rho r)$ \citep{wil83} and assume $\rho=300~\mathrm{kg \cdot m^{-3}}$, $\beta=0.01$ is equivalent to $r=200~\mathrm{\mu m}$ or $10^{-9}~\mathrm{kg}$ for a spherical particle. This number is significant because $10^{-9}~\mathrm{kg}$ is considered to be the lower end of threat regime by spacecraft designers \citep{mcn04}. Since optical observation strongly favors micron-sized particles due to their higher scattering efficiency, the absence of tail structure at larger $\beta$ suggests that the cometary tail is dominated by larger, spacecraft-threatening particles.

\section{The Dust Tail Model}

\subsection{Philosophy of the Model}

We develop a semi-analytic dust tail model (DTM) to parameterize the cometary dust tail. The key of parameterization involves the initial velocity, i.e. the velocity vector of the particles at the time it is ejected from the cometary nucleus. We start from the revised \citet{whi51}'s model by \citet{bro98} which was used to study the Perseid meteor stream formed by 109P/Swift-Tuttle:

\begin{equation}
V = (10.2~\mathrm{m \cdot s^{-1}})~\left( \frac{r_h}{1~\mathrm{AU}} \right)^{-\frac{1}{2}}~\left( \frac{\rho}{1~\mathrm{g \cdot cm^3}}\right)^{-\frac{1}{3}}~\left( \frac{r_{C}}{1~\mathrm{km}}\right)^{\frac{1}{2}}~\left(\frac{m}{1~\mathrm{g}}\right)^{-\frac{1}{6}}
\label{eqn-bro98}
\end{equation}

where $V$ is the ejection velocity relative to the nucleus in $\mathrm{m \cdot s^{-1}}$, $r_h$ is heliocentric distance of the comet, $\rho$ is the bulk density of the meteoroids, $r_{C}$ is the radius of the cometary nucleus, and $m$ is the mass of the meteoroid.

Brown \& Jones reported a satisfactory agreement between the model and the observation for the case of Perseid meteor stream, however they noted that the agreement might be due to the nature of high ejection velocity of P/Swift-Tuttle and may not be applicable to other comets. To accommodate this issue, we introduce a ``reference'' ejection velocity that had been used in other studies \citep[such as][]{ish07} and rearrange the terms:

\begin{equation}
V = V_0~\left( \frac{r_h}{1~\mathrm{AU}} \right)^{-u}~\left( \frac{a}{a_0} \right)^{-\frac{1}{2}}~\left( \frac{\rho}{\rho_{0}} \right)^{-\frac{1}{2}}~\left( \frac{d_C}{1~\mathrm{km}} \right)^{\frac{1}{2}}
\label{eq-v}
\end{equation}

where $V_0$ (in $\mathrm{m \cdot s^{-1}}$) is a reference ejection velocity of particles with \textit{size} $a_0=5~\mathrm{mm}$ and bulk density $\rho_0=1000~\mathrm{kg \cdot m^{-3}}$ ejected by a cometary body of \textit{diameter} of 1~km at a heliocentric distance of $1~\mathrm{AU}$, and $u$ is the dependence on heliocentric distance; the introduction of $u$ will be elaborated in the next section. The $V_0$ corresponding to the constant of 10.2 in Brown \& Jones' model would be $V_0=8.0~\mathrm{m \cdot s^{-1}}$. We assume the particles are symmetrically released at the comet's sub-solar point at a direction $w<45^{\circ}$ from the sunward direction. This $w$ limit is chosen following the results of \citet{ish07} and \citet{ish08}.

The initial velocity model is fed by a Monto-Carlo subroutine that generates random particle with sizes following a power law: $N(a)=\left( \frac{a}{a_0} \right) ^{q}$, where $N(a)$ is the accumulative particle number, and $q$ is the size distribution index. We then solve Kepler's equation rigorously from the start time to the time of observation to determine the position of the particle. These steps are repeated until we have a sufficient number of particles to simulate the morphology of the cometary tail.

At the end, we compute the spatial intensity on a sky plane coordinate ($\alpha$, $\delta$). We consider a simple model without secondary effects such as the response efficiency of CCD sensor to different wavelength and the scattering efficiency due to particle shape. We consider the light contribution from each particle:

\begin{equation}
I'(a,r_h) = \left( \frac{r_h}{1~\mathrm{AU}} \right)^{-2} a^2 A_p
\end{equation}

where $A_p$ is the modified geometric albedo. The intensity at ($\alpha$, $\delta$) will just be the sum of intensities from all particles within the region ($d \alpha' d \delta'$):

\begin{equation}
I(\alpha, \delta) = \iint I'(a,r_h) d \alpha' d \delta'
\end{equation}

Finally, a 2-dimensional intensity map is created by looping around all possible ($\alpha$, $\delta$) and computing the value of $I(\alpha, \delta)$ at each position.

\subsection{Determining the $u$ Constant}

The treatment of $u$ is somewhat tricky, as a number of $u$ have been suggested by previous workers. For example, some studies from both cometary and meteor communities suggested $u=0.5$ \citep[e.g.][]{cri95,bro98,ish07,ish08}, \citet{whi51}'s original model suggested $u=1.125$, while $u=1$ \citep{bro98,rea00,ma02} and $u=3$ \citep{aga10} were also used. We notice that most studies adopted $u=0.5$ do not have data that cover a broad range of $r_h$; our initial test with $u=0.5$ also shows noticeable mismatch at large $r_h$ (e.g. $r_h>3$~AU). This may be due to the fact that the range of $r_h$ is not broad enough to constrain $u$ more effectively, but could also due to, for example, the onset of water-ice sublimation at $\sim2.3$~AU that dramatically enhances the ejection regime and hence no unique $u$ can be defined.

To investigate this matter, we use the observations of C/2012 S1 (ISON) which are available in a broad range of $r_h$. The unfiltered observations were taken at Xingming Observatory, China, from a few days after the discovery ($r_h=6.2$~AU) to a few weeks before the perihelion ($r_h\lesssim 1$~AU, see Table~\ref{tbl-obs-ison}). The DTM model is run at grids with $u\in[0.5, 4.0]$ and $V_0\in[1.0, 8.4]~\mathrm{m \cdot s^{-1}}$, with orbital elements from JPL~54 (Table~\ref{tbl-orb}). A few parameters at the far ends are not tested as initial trials indicate that they are unlikely to contain the best fits.

Except the nucleus size of C/ISON, which has been reported to be at the order of 1~km \citep{kni13,li13}, we have to make some assumptions, such as the albedo and particle density. Although the nucleus size will not effectively affect our final result (since its contribution is modest in most cases and can be balanced by a slightly larger or smaller $V_0$ term), we still keep it in the simulation as we hope that $V_0$ can be comparable to C/Siding Spring and other comets. The set of parameters are summarized in Table~\ref{tbl-dtm-ison}. The simulation results and the observations are then compared and graded separately by both authors on a Boolean basis (i.e. as ``possible fit'' or ``definitely not a possible fit''). Finally, the grades are summed and is scale to a score from 0 to 100.

\begin{figure*}
\includegraphics[width=\textwidth]{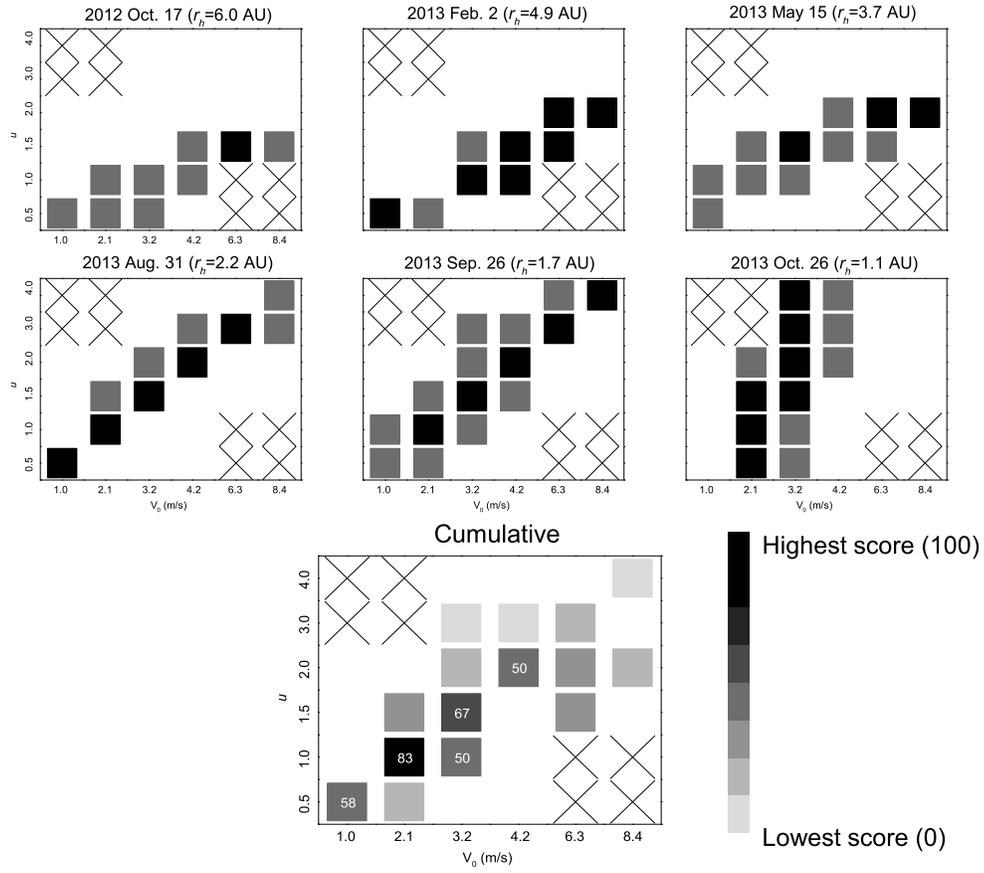}
\caption{The score charts for C/ISON for determining the $u$ constant. Cross mark indicates the trials that are not evaluated. Scores (between 0 and 100, with 100 to be the highest) are listed if $>50$.}
\label{fig-ison-score}
\end{figure*}

The final score chart (Figure~\ref{fig-ison-score}) indicates that $V_0=2.1$~(m/s), $u=1.0$ is the best fit. It is encouraging that no dramatic morphological change is presented near the water-ice sublimation line ($\sim2.3$~AU), which means that the $u$ we found is an unique approximation to the entire $r_h$ range. The change at $r_h=1.1$~AU case is most likely due to the contamination from cometary gas emission, as no filters were used to block the primary emission lines (e.g. C$_2$ and C$_3$ that falls in the $\lambda=500$~nm range where the imaging CCD is sensitive at); we may remove the $r_h=1.1$~AU case and it does not alter our result.

\subsection{Simulation Result}

After pinned down the $u$ constant, we run the DTM model for the case of C/Siding Spring with $V_0$ ranging from 1.0 to 4.2~m/s to determine $V_0$. The input parameters and orbital elements summarized in Table~\ref{tbl-orb} and \ref{tbl-dtm-para}. Since the epochs of our observations are fairy close, we only simulate the Nov. 23 case, as the master frames were stacked from more frames and had a slightly better airmass. We estimate $d_C=5$~km considering two comparable comets which the nucleus diameters are constrained at a higher level of confidence: 19P/Borrelly (M1=8.9, M2=13.8, $d_C=4.8$~km)\footnote{\url{http://ssd.jpl.nasa.gov/sbdb.cgi?sstr=19P}, retrieved 2014 Mar. 2.} and C/1996 B2 (Hyakutake) (M1=7.3, M2=11.1, $d_C=4.2$~km)\footnote{\url{http://ssd.jpl.nasa.gov/sbdb.cgi?sstr=1996+B2}, retrieved 2014 Mar. 2.}, in contrast C/Siding Spring has M1=8.6, M2=10.4.

\begin{figure}
\includegraphics[width=0.5\textwidth]{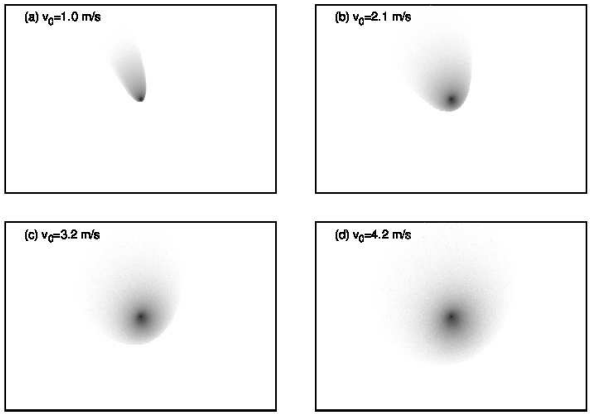}
\caption{DTM simulation for the master frame of C/Siding Spring on 2013 Nov. 23. The size of each figure is 8'$\times$4'. North is up and east is left.}
\label{fig-a1dtm}
\end{figure}

The simulation result is shown in Figure~\ref{fig-a1dtm}. Comparing the simulation to the observation (Figure~\ref{fig-img}), we see $V_0=1.0$~m/s as the optimal match. We note that the strong size gradient indicated in the syndyne diagram (Figure~\ref{fig-syn}) actually forged the result: since optical observation strongly favors smaller particles that dominates the west-end sector of the tail (i.e. toward the clockwise direction), the west-end boundary (about P.A.~$\sim345^{\circ}$) marks the true boundary of the tail. As we only simulate particles with $\beta<0.01$, the simulated particles should mostly distribute within the anticlockwise direction of P.A.~$\sim345^{\circ}$, which only the $V_0=1.0$~m/s case satisfies. Such converge allows us to make a stronger conclusion on the optimal $V_0$.

The $V_0=1.0$~m/s value is comparable to some comets that were previously studied, including 4P/Faye, 22P/Kopff and the Rosetta mission target, 67P/Churyumov-Gerasimenko \citep[c.f.][note that the $V_0$ here needs to be $\times4$ to make the numbers comparable]{rya13}, although the dynamical origins of these comets seems to be different. It is interesting to note that this similarity comes with the fact that comets with similar dynamical origins can have a much deviated $V_0$: for example, $V_0$ for 2P/Encke is about a magnitude larger. This may be purely by coincidence, but further studies with larger sample size may determine if there is anything physical.

\section{Encounter}

Eventually, we run the simulation for the encounter with best-fit parameters found in the previous iterations (summarized in Table~\ref{tbl-dtm-para}), to study the influence of cometary dust at Mars and its vicinity. Unlike other meteor stream models, our model does not include planetary perturbation; but we think this is acceptable for the case of C/Siding Spring since the comet is far from the ecliptic plane ($i=129^{\circ}$) until the encounter.

The synchronic feature on Figure~\ref{fig-syn} suggests that C/Siding Spring was active at least a year before our observations, therefore we choose $\tau_{\mathrm{max}}=1000$~d (corresponding to 2012 Jan. 23) to be the start date of the simulated cometary activity to catch any early activities. State vectors of simulated particles are recorded within $\pm 1$~d of the closest encounter (2014 Oct. 18, 18:30 UT to 2014 Oct. 20, 18:30 UT) in a step of 1~min. Since simulating the entire set of cometary particles will take $quite$ a bit of time (comets typically release particles at a rate of $\sim10^{7}-10^{10}~\mathrm{s^{-1}}$), we only simulate a representative number of particles and scale them up afterwards.

We first need to examine the dust production rate of C/Siding Spring; this can be tied to the so-called $Af\rho$ quantity \citep{ahe84}. The number of ejected particles in the particle size range ($a_1$,$a_2$) can be related to $Af\rho$ by \citep{vau05,ye14}:

\begin{equation}
Q_g(a_1,a_2)=\frac{655 A_1(a_1,a_2) Af\rho}{8 \pi A_B j(\phi)[A_3(a_1,a_2)+1000A_{3.5}(a_1,a_2)]}
\end{equation}

where $A_x=(a_2^{x-s}-a_1^{x-s})/(x-s)$ for $x\neq s$ and $A_x=\ln(a_2/a_1)$ for $x=s$, with $s$ is the size population index, $A_B$ is the Bond albedo and $j(\phi)$ is the normalized phase function.

We obtain the $Af\rho$ measurements conducted by a group of observers and collected by the ``Cometas Obs''\footnote{Available at \url{http://www.astrosurf.com/cometas-obs/C2013A1/afrho.htm}, retrieved on 2014 Feb. 6.}. The measurements show a steady $Af\rho$ near 1500~cm from $r_h=6.76$~AU to $r_h=3.72$~AU. By assuming $Af\rho \propto \left( \frac{q}{r_h} \right)^{\frac{4}{3}}$ when the comet gets into the water-ice sublimation line (to include possible early onsets, we use a loose constraint, $r_h\simeq3$~AU), we estimate the $Af\rho$ of C/Siding Spring at 1~AU to be 3700~cm, which corresponds to $N_0 = 3\times10^{10}~\mathrm{s^{-1}}$.

\begin{figure}
\includegraphics[width=0.5\textwidth]{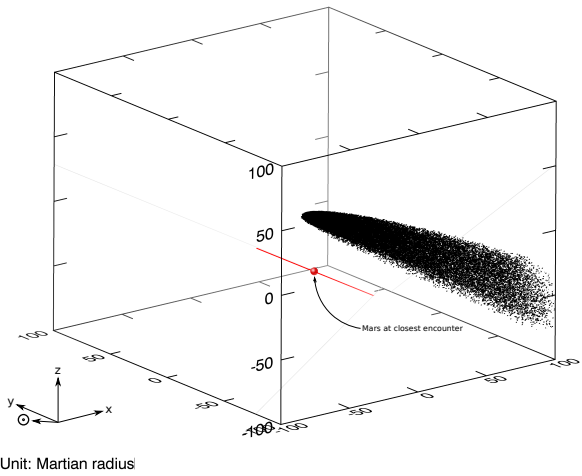}
\caption{The DTM simulation result together with the trajectory of Mars shown in a cometocentric ecliptic frame.}
\label{fig-plot2r}
\end{figure}

The result is shown in Figure~\ref{fig-plot2r}: Mars will miss the dust cone by some 20 Martian radius or 67,800~km.

\begin{figure}
\includegraphics[width=0.5\textwidth]{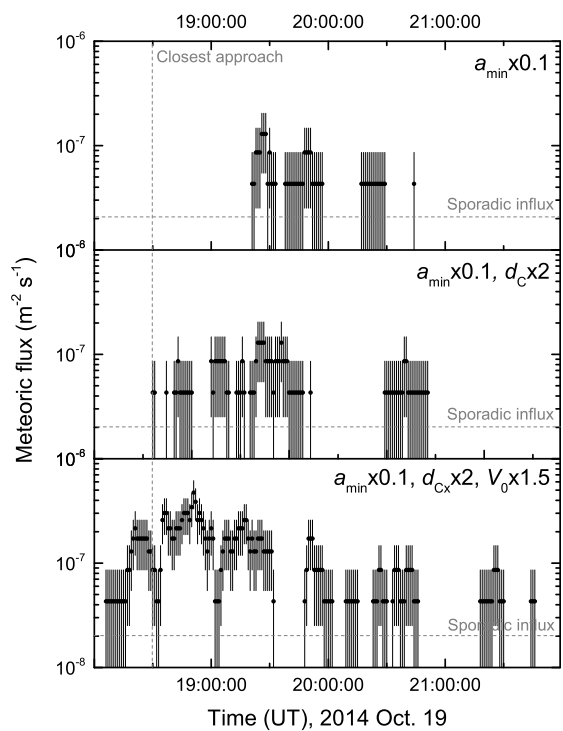}
\caption{The meteoroid influx at Mars at three scenarios (see main text). Uncertainty bar depicts the Poisson error ($\sigma=\sqrt{N}$). The ``peak-lets'' of fluxes are due to low statistics.}
\label{fig-flux}
\end{figure}

What about the extreme cases in the uncertainty ranges? To investigate this, we run further simulations with some educative guesses about the uncertainty ranges: a factor of 10 for the minimum particle size, a factor of 2 for the diameter of the cometary nucleus, and 50\% for the reference velocity. Four combinations are tested with other parameters remain the same as Table~\ref{tbl-dtm-para}. The combinations and results are shown in Table~\ref{tbl-flux} and Figure~\ref{fig-flux}.

For scenario 4, which the minimum particle size/mass remains unchanged (i.e. particles are confined within the spacecraft-threatening category), we still find no direct encounter between Mars and the dust cone. For the other three scenarios, the dust cone does reach Mars. The peak times are about 30--60~min. behind the closest approach. We see some ``peak-lets'' in the time series plots which should be artifacts due to low statistics rather than anything physical. The low statistics also make it difficult to determine the duration of the event, but we can crudely estimate the Full-Wide-Half-Maximum to be $\sim1$~hr or less. The peak fluxes are at the order of $10^{-7}~\mathrm{m^{-2} \cdot s^{-1}}$ while the accumulative fluxes are at the order of $10^{-5}~\mathrm{m^{-2}}$, appropriate to the meteoroids larger than $10^{-12}$~kg. The absence of encounter in scenario 4 suggests that the particles that arrive Mars in scenarios 1--3 are at the range of $10^{-12}$ to $10^{-9}$~kg, which are below the spacecraft-
threatening regime.

\section{Discussion}

\subsection{Comparison with \citet{moo14}}

\citet{moo14} also studied the meteoroid influence at Mars during the encounter of C/Siding Spring but with a different approach: they construct a symmetric analytic model of the coma that relates the dust production rate to the total magnitude (M1) of the comet. As a separate check, the meteoroid flux experienced by Giotto and Stardust spacecrafts during their encounter with 1P/Halley and 81P/Wild were reproduced by this model and compared with the actual data, which order-of-magnitude agreement were found. Since more parameters were on loose at the time of their study, they parameterized their result and reported an accumulative flux of $0.15~\mathrm{m^{-2}}$ for particles over $4.19\times10^{-9}$~kg under nominal conditions. They also conducted simulation with a dynamical meteor stream model that uses \citet{bro98}'s ejection model (Wiegert, private comm.) and found the numbers from the two models agreed within an order of magnitude, although the number from the dynamic model is in fact lower by a factor 
of 2, of which they attributed to the difference in modeling assumptions.

Since their study, the M1 of C/Siding Spring has been revised from 5.2 to 8.6; using Eq.~18 in their study and $\rho=300~\mathrm{kg \cdot m^{-3}}$, we find a new value of $0.0135~\mathrm{m^{-2}}$. This is still more than three order of magnitude higher than our result. To investigate this difference, we test our model with $V_0$ set to the value equivalent to the \citet{bro98}'s model and find a much closer value ($\sim0.001~\mathrm{m^{-2}}$), which would allow an order-of-magnitude agreement to Moorhead et al's result from the dynamical model. From here we suggest that the difference between the two results is primary contributed by different input parameters.

\subsection{Implication to the Spacecrafts, Rovers and Martian Moons}

The sporadic meteoroid influx on Mars is about half of the influx on Earth \citep{dom07} or $\sim1.8\times10^{-8}~\mathrm{m^{-2} \cdot s^{-1}}$ for meteoroid larger than $10^{-9}$~kg. For comparison, the peak flux of the 2012 Draconid meteor storm is $3.3\times10^{-6}~\mathrm{m^{-2} \cdot s^{-1}}$ for meteoroid larger than $10^{-9}$~kg assuming a general power law distribution \citep{ye14b}\footnote{The power law distribution is quoted from \citet[][Table XXVI]{cep98}, which suggested a factor of $10^{3.69}$ between the cumulative number of meteoroids larger than $10^{-7}$ and $10^{-9}$~kg respectively.}. From here, our simulation seems to indicate that the meteoric influx at Mars due to C/Siding Spring is comparable to the sporadic background. For the case of smaller-sized meteoroids, the sporadic influx is about 3 magnitudes higher from $10^{-9}$~kg to $10^{-12}$~kg, which would still put the influx due to C/Siding Spring no higher than the sporadic background. The simulation also suggested that the bulk 
of the outburst (if any) would not be longer than the orbital period of the orbiters around Mars. For the Opportunity and Curiosity rovers, the potential meteor outburst takes place at 
Martian morning and afternoon respectively, which prevent any meteor observations from the rover.

The impact on Martian moons is another interesting topic. At the time of the encounter, Deimos will be about 2 Martian radius closer to the cometary nucleus than Mars itself. Depending on its physical property and impact angle, a $200~\mathrm{\mu m}$ meteoroid may produce a sub-meter crater on Phobos and Deimos\footnote{Estimate with \url{http://www.lpl.arizona.edu/tekton/crater_c.html}.}. Unfortunately, such crater is below the resolution of our current best images (about 5~m per pixel), but it is worth pointing out that the high speed nature of meteoroids originated from C/Siding Spring should create larger craters more efficiently than nominal sporadic meteoroids.

\subsection{Possibility of Gravitational Disrupted Tail}

\begin{figure*}
\includegraphics[width=\textwidth]{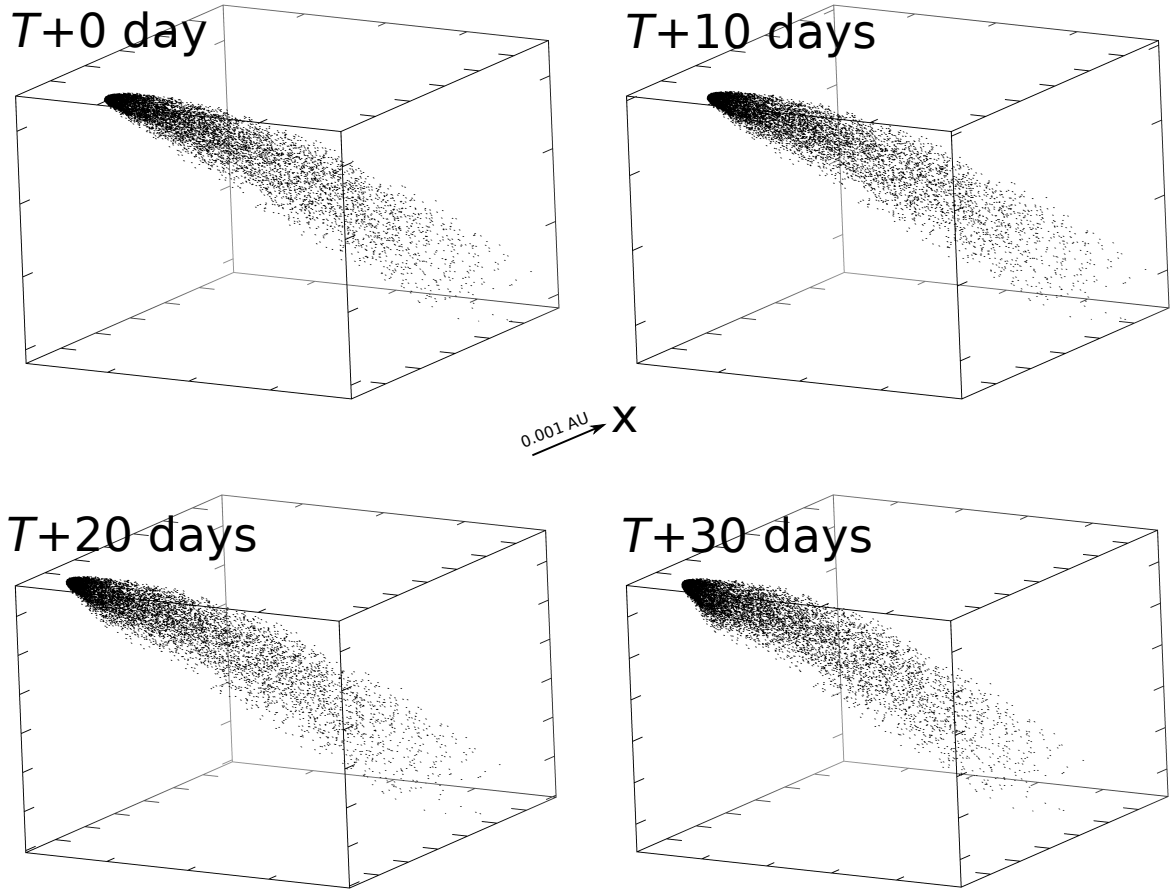}
\caption{The evolution of dust tail after the gravitational perturbation of Mars, from $T+0$~day (encounter) to $T+30$~days. The graphs are constructed in ecliptic coordinate system, with X-axis pointing to the vernal point.}
\label{fig-pert3d}
\end{figure*}

\begin{figure*}
\includegraphics[width=\textwidth]{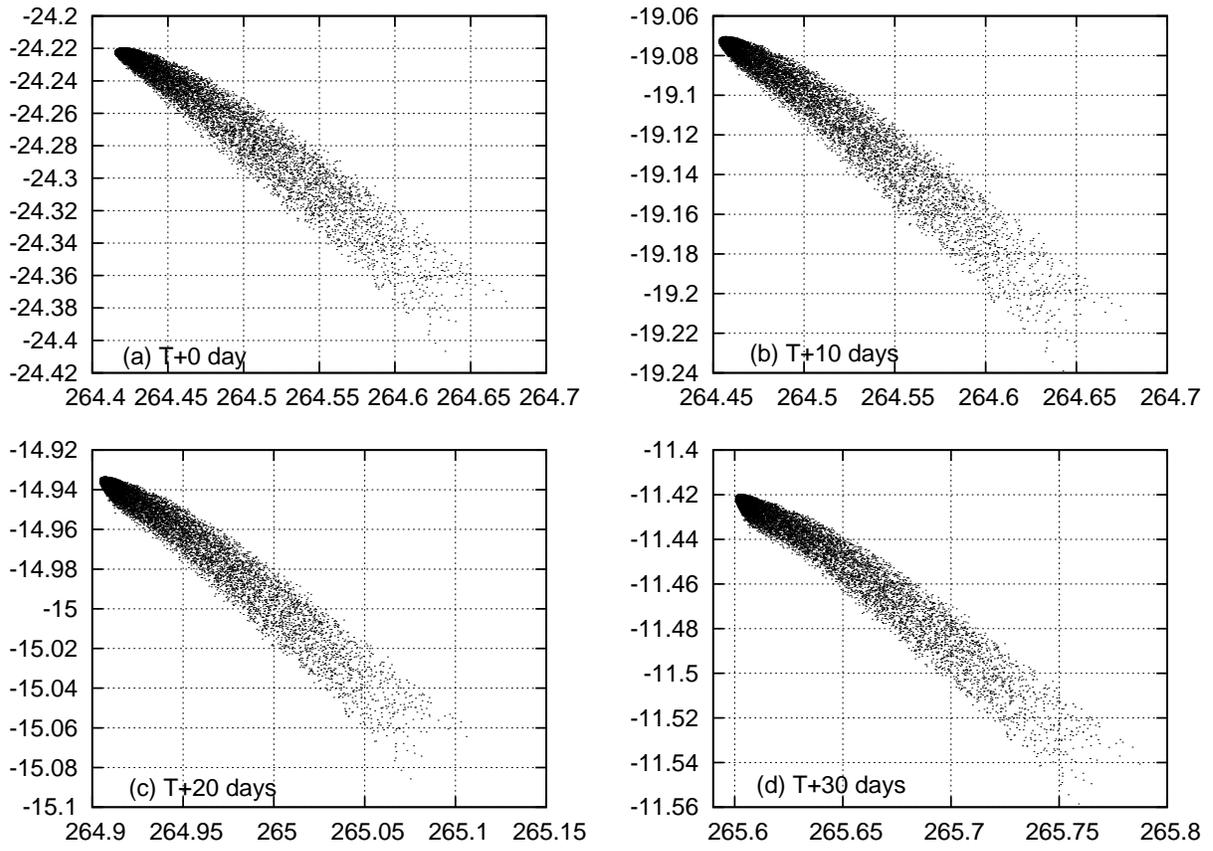}
\caption{The evolution of dust tail after the encounter as observed from the Earth.}
\label{fig-pertobs}
\end{figure*}

As one of the closest cometary approaches to a major planet among known objects, we expect the dust tail of C/Siding Spring to be disrupted by the gravitational field of Mars to some degree. To investigate this matter, we integrate a snapshot of the locations of some 12,000 dust particles (generated from previous simulations under the nominal condition) from $T+0$ to $T+30$ days (with $T$ being the time of closest approach). The integration is performed with the HNBODY package \citep{rau02} using the symplectic intergrator, with the barycenter of the Martian system included. The result is shown as Figure~\ref{fig-pert3d} and \ref{fig-pertobs}. We find that at $T+20$~days, the apparent size of the ``clump'' reaches the order of $0.01^{\circ}$ or 30'' as seen on Earth, which may be detectable by ground-based telescopes.

\section{Conclusion}

We reported the observations and modeling works of C/Siding Spring, a dynamically new comet that will make a close approach to Mars on 2014 Oct. 19. By fitting the observations with syndyne simulations, we found that the tail of C/Siding Spring was dominated by larger particles at the time of observation. Synchrone simulation suggested that the particles dominate the optical tail was released by the comet as early as late 2012, when the comet was more than $\sim7$~AU from the Sun. We then developed a semi-analytic model to simulate the cometary dust activity. The modeling result suggested a modest ejection velocity of C/Siding Spring that is comparable to a few other comets, including P/Churyumov-Gerasimenko, target of the Rosetta mission.

The same model was then used to study the meteoroid influence to Mars during the encounter, fed with the constraints found in the previous steps. We found that the planet will miss the dust cone by some 20 Martian radius (67,800~km) under nominal situation. Although the planet may be engulfed by the cometary dust tail if we made an educative guess about the uncertainties and pushed the parameters to the extreme cases, the simulation suggested that the meteoroids reach the vicinity of Mars are dominated by non-spacecraft-threatening meteoroids, and the meteoric influx is not significantly higher than the sporadic background influx; the duration of the event is at the order of 1~hr. From our simulation, it seems that intense and enduring meteoroid bombardment at Mars and its vicinity region is unlikely during the flyby of C/Siding Spring.

We also study the potential gravitational disruption of the cometary dust tail. A simple numerical integration suggested that the dust ``clump'' created by the gravitational drag would be at the order of tens of arcsecs at $T+20$~days as seen from the Earth which may be detectable by ground-based facilities.

At the time of the writing, C/Siding Spring is about 4~AU from the Sun. As the comet travels into the inner solar system and enters the water-ice sublimation line, the story could evolve dramatically. We encourage observers to closely monitor C/Siding Spring as it helps on creating a full picture of this unprecedented cosmic event.

\textit{Note:} at the reviewing stage of this paper, J. Vaubaillon et al. also reported their modeling result of the same event \citep[see][]{vau14}. Our initial check using their input values suggested an order-of-magnitude agreement between the two results, which indicated that the difference between the two results is primary due to input parameters.

\acknowledgments

We thank an anonymous referee for his/her comments. We are indebted to the Jade Scope team led by Angus Lau, who kindly agreed us to use their facility in Australia and did a lot of work on executing the observations and dealing with instrumental issues. We also thank Xing Gao of Xingming Observatory for the C/ISON data and his long-term support on the Xingming facilities. We thank Peter Brown, Matthew Knight, Jian-Yang Li and Paul Wiegert for their helps and discussions. Q.-Z. thanks Summer Xia Han for help fighting his laziness and urging him to finish up the manuscript.

We also thank the following ``Cometas Obs'' observers for their $Af\rho$ measurements: Jos\'{e} Ram\'{o}n Vidal Blanco, Josep Mar\'{\i}a Bosch, Montse Campas, Garc\'{\i}a Cuesta, Jos\'{e} Francisco Hern\'{a}ndez, Gustavo Muler, Ramon Naves, and the unspecified observers at Moorook and Siding Spring Observatories.


\clearpage

\clearpage

\clearpage

\clearpage

\clearpage

\clearpage

\clearpage

\clearpage

\clearpage

\clearpage

\clearpage

\clearpage

\clearpage

\begin{table*}
\begin{center}
\caption{Observation circumstances of the C/Siding Spring data. \label{tbl-obs}}
\begin{tabular}{lccccc}
\hline
Date (UT) & Exposure & Total frames & Airmass & $r_h$ & $\Delta$ \\
          & & & & (AU) & (AU) \\
\hline
2013 Nov. 12.63 & 90~s & 34 & 1.010 & 4.450 & 3.857 \\
2013 Nov. 20.56 & 90~s & 60 & 1.019 & 4.374 & 3.776 \\
2013 Nov. 23.59 & 90~s & 63 & 1.012 & 4.345 & 3.750 \\
\hline
\end{tabular}
\end{center}
\end{table*}

\clearpage

\begin{table*}
\begin{center}
\caption{Observation circumstances of the C/ISON data. \label{tbl-obs-ison}}
\begin{tabular}{llccccc}
\hline
Date (UT) & Instrument & Pixel size & Exposure & Total frames & $r_h$ & $\Delta$ \\
          & & per pixel & & & (AU) & (AU) \\
\hline
2012 Oct. 17.90 & 0.35-m SASP & 0.9'' & 90~s & 50 & 6.029 & 6.051 \\
2013 Feb. 2.69 & 0.35-m SASP & 0.9'' & 90~s & 60 & 4.916 & 4.026 \\
2013 May 15.64 & 0.35-m SASP & 0.9'' & 60~s & 34 & 3.710 & 4.342 \\
2013 Aug. 31.92 & 0.11-m XP1 & 3.5'' & 60~s & 10 & 2.179 & 2.973 \\
2013 Sep. 26.91 & 0.11-m CSP & 3.5'' & 120~s & 30 & 1.726 & 2.667 \\
2013 Oct. 26.94 & 0.35-m SASP & 0.9'' & 20~s & 20 & 1.116 & 1.374 \\
\hline
\end{tabular}
\end{center}
\end{table*}

\clearpage

\begin{table}
\begin{center}
\caption{Orbital elements (in J2000) of C/ISON (from JPL 54) and C/Siding Spring (from JPL 40) used for the simulation. \label{tbl-orb}}
\begin{tabular}{lcc}
\hline
Element & C/ISON & C/Siding Spring \\
\hline
$q$ [AU] & 0.01245259 & 1.3991521\\
$e$ & 1.000201 & 1.00057866\\
$i$ & $62.4039775^{\circ}$ & $129.025859^{\circ}$\\
$\Omega$ & $295.6520316^{\circ}$ & $300.972350^{\circ}$\\
$\omega$ & $345.5312406^{\circ}$ & $2.432872^{\circ}$\\
$t_p$ (UT) & 2013 Nov. 28.77866 & 2014 Oct. 25.40490\\
\hline
\end{tabular}
\end{center}
\end{table}

\clearpage

\begin{table*}
\begin{center}
\caption{DTM parameters for the simulation of C/ISON. \label{tbl-dtm-ison}}
\begin{tabular}{lcc}
\hline
Parameter & & Value \\
\hline
Maximum $\tau$ [d] & $\tau_{\mathrm{max}}$ & 365 \\
Particle size [m] & $a$ & [$10^{-5}$, $10^{-2}$] \\
Size distribution index & $s$ & -2.6 \\
Modified geometric albedo & $A_p$ & 0.04 \\
Number-distance constant & $k$ & -3.0 \\
Bulk density of particle [$\mathrm{kg m^{-3}}$] & $\rho$ & 300 \\
Size of cometary nucleus [km] & $d_C$ & 1 \\
Dependence constant on heliocentric distance & $u$ & 0.5, 1.0, 1.5, 2.0, 3.0, 4.0 \\
\hline
\end{tabular}
\end{center}
\end{table*}

\clearpage

\begin{table*}
\begin{center}
\caption{DTM parameters for the simulation of C/Siding Spring. \label{tbl-dtm-para}}
\begin{tabular}{lcc}
\hline
Parameter & & Value \\
\hline
Maximum $\tau$ [d] & $\tau_{\mathrm{max}}$ & 1000 \\
Particle size [m] & $a$ & [$10^{-4}$, $10^{-2}$] \\
Size distribution index & $s$ & -2.6 \\
Modified geometric albedo & $A_p$ & 0.04 \\
Number-distance constant & $k$ & -3.0 \\
Bulk density of particle [$\mathrm{kg m^{-3}}$] & $\rho$ & 300 \\
Size of cometary nucleus [km] & $d_C$ & 5 \\
Dependence constant on heliocentric distance & $u$ & 1.0 \\
\hline
\end{tabular}
\end{center}
\end{table*}

\clearpage

\begin{table*}
\begin{center}
\caption{The meteoroid influence at Mars under four extreme cases in the uncertainty ranges. \label{tbl-flux}}
\begin{tabular}{llllll}
\hline
Scenario & 1 & 2 & 3 & 4 \\
Condition & $a_{\mathrm{min}} \times 0.1$ & $a_{\mathrm{min}} \times 0.1$ & $a_{\mathrm{min}} \times 0.1$ & $d_C \times 2$ \\
 & & $d_C \times 2$ & $d_C \times 2$ & $V_0 \times 1.5$  \\
 & & & $V_0 \times 1.5$ \\
\hline
Active time (approx.) & 19:20--20:45 & 18:30--20:55 & 18:05--21:50 & N/A \\
 (UT, 2014 Oct. 19) & & & & \\
Peak time and flux & 19:27 & 19:26 & 18:51 & N/A \\
 (UT, 2014 Oct. 19) & & & \\
Peak flux & $1.3\times10^{-7}$ & $1.3\times10^{-7}$ & $4.7\times10^{-7}$ & N/A \\
 ($\mathrm{m^{-2} \cdot s^{-1}}$) & & & & \\
FWHM & $<$1~hr. & $<$1~hr. & 1~hr. & N/A \\
Total flux ($\mathrm{m^{-2}}$) & $2.6\times10^{-6}$ & $4.8\times10^{-6}$ & $1.8\times10^{-5}$ & N/A \\
\hline
\end{tabular}
\end{center}
\end{table*}

\end{CJK*}
\end{document}